\newcommand{\shorttitle}{Quantum Kinetic Theory for Plasmas in Strong Fields}
\newcommand{\shortauthor}{M. Bonitz, Th. Bornath, D. Kremp}
\newcommand{\contnum}{329} 
\newcommand{\contvolum}{39} 
\newcommand{\contyear}{1999} 
\newcommand{\contpages}{} 
\newcommand{\be}{\begin{equation}}
\newcommand{\ee}{\end{equation}}

\documentstyle[twoside,amssymb]{cpp}

\begin{document}
\setcounter{page}{1}
\title{
Quantum Kinetic Theory for Laser Plasmas.\\
Dynamical Screening
in Strong Fields
}

\author{
M. Bonitz$^{\rm(a)}$, Th. Bornath$^{\rm(a)}$, D. Kremp$^{\rm(a)}$,
M. Schlanges$^{\rm(b)}$, and W.D. Kraeft$^{\rm(b)}$
}

\address{
(a) Universit\"at Rostock, Fachbereich Physik,
Universit\"atsplatz 3, D-18051 Rostock \\
(b) Ernst-Moritz-Arndt Universit\"{a}t Greifswald,
Institut f\"{u}r Physik, \\Domstr.~10a, 17487 Greifswald, Germany
}

\maketitle

\begin{abstract}
A quantum kinetic theory for correlated charged--particle systems in
strong time--dependent electromagnetic fields is developed. Our approach is
based on a systematic gauge--invariant nonequilibrium Green's functions
formulation. Extending our previous analysis \cite{kremp-etal.99pre}
we concentrate on the selfconsistent treatment of dynamical screening
and electromagnetic fields which is applicable to arbitrary nonequilibrium
situations.
The resulting kinetic equation generalizes previous
results to quantum plasmas with full dynamical screening and includes
many--body effects. It is, in particular, applicable to the interaction
of dense plasmas with strong electromagnetic fields, including laser fields
and x-rays. Furthermore, results for the modification of the plasma screening
and the longitudinal field fluctuations due to the electromagnetic
field are presented.
\end{abstract}
\section{Introduction}

With the progress in short--pulse laser technology \cite{perry-etal.94}
high intensity electromagnetic fields are becoming broadly available.
In particular, they make it possible to create strongly
correlated quantum plasmas under extreme nonequilibrium conditions which
opens a broad range of applications, e.g. \cite{Tea94}.
At the same time, optical techniques for
time--resolved diagnostics are improving remarkably \cite{theobald-etal.96}.
These developments create the need for a quantum kinetic theory of dense
nonideal plasmas in intense laser fields.

Nonequilibrium properties of dense plasmas in which collisions are important
are usually studied on the basis of
kinetic equations of the Boltzmann type.
However, in spite of their fundamental character, Boltzmann--like kinetic
equations have a number of shortcomings, in particular in view of their application
to dense plasmas in intense laser fields:
\begin{description}
\item[i)]{they are valid only for times larger than the correlation (or
collision) time $\tau_{{\rm corr}}\sim \omega_{pl}^{-1}$,}
\item[ii)]{they conserve only the mean kinetic energy
instead of the sum of kinetic and potential energy,}
\item[iii)]{they are valid only in the weak field limit since the
corresponding collision integrals are independent of the
electromagnetic field,}
\item[iv)]{they are not applicable to high--frequency processes (fields),
where $\omega > \omega_{pl}$, cf. i).}
\end{description}
Obviously, in the case of strong correlations, high--frequency
electromagnetic fields and/or short--time phenomena generalizations are
necessary.

{\em Generalized kinetic equations for correlated plasmas} have been derived
already in the 60ies by Prigogine \cite{prigogine}, Zwanzig \cite{zwanzig},
Kadanoff and Baym \cite{kadanoff-baym,BK61}, Balescu \cite{balescu63,balescu},
Silin \cite{silin67}, Klimontovich
\cite{klimontovich64,KE72,klimontovich75} and others. In recent years,
the increasing interest in ultrafast processes has revived the theoretical
activities, e.g. \cite{BK96,KBKS97,KBBS96,BKKS96}, acompanied by
progress in numerical solutions, e.g. \cite{BK96,BKS+96,banyai-etal.98};
for textbook overviews, see \cite{haug-jauho96,bonitz-book}.
Furthermore, kinetic equations for classical plasmas in
{\em high--frequency fields} have been derived for the first time in
papers of Silin \cite{silin60,silin61,silin64}. Among other problems,
he computed
the high--frequency conductivity of a plasma. We mention that, for the
weak--field limit, this problem has been studied by many authors,
including Oberman et al. \cite{dawson-etal.62,oberman-etal.62}
and DuBois et al.
\cite{dubois-etal.63}. An essential further development of the
theory has been given by Klimontovich and co--workers
\cite{KP74,klimontovich75}. Klimontovich used his powerful technique of
second quantization in phase space \cite{Kli57,klimontovich64} to investigate
the density--density and microfield fluctuations in low and high--frequency
fields. This allowed him to derive collision integrals for classical plasmas
in strong fields which take into account dynamical screening and to derive
a complete theory of transport processes \cite{KP74,klimontovich82}.
A result of central importance is an expression for the collisional
heating rate and the electron--ion collision frequency in strong fields
in terms of the imaginary part of the inverse dielectric function
${\rm Im}\epsilon^{-1}$ \cite{klimontovich75}. Recently, expressions of the
same form were derived again \cite{decker-etal.94}. For a recent
overview on the collision frequency in laser plasmas, we refer to Mulser
et al. \cite{mulser-etal.97,mulser-etal.99}.

The above kinetic theories for plasmas in electromagnetic fields were
limited to classical plasmas. A first extension to quantum plasmas was
given by Silin and Uryupin \cite{silin-ur}.
More recently, a kinetic equation for dense quantum plasmas
in strong static fields has been derived \cite{morawetz-etal.93,msk},
whereas a systematic quantum kinetic theory for plasmas in strong fields of
arbitrary time dependence was presented in ref. \cite{kremp-etal.99pre}.
There, electron--electron and electron--ion quantum collision integrals
were derived within the static Born approximation (Landau collision
integrals).

In this paper, we extend this theory to the case of
full dynamical screening for dense quantum plasmas under arbitrary
nonequilibrium conditions by using the random phase approximation
(polarization approximation) for the particle--particle scattering
processes. Our approach is based on the nonequilibrium Green's functions
formalism which allows for the most straightforward derivation and for an
explicit solution of the gauge problem.
We derive a kinetic equation which is a generalization of
Klimontovich's classical result \cite{KP74} and, on the other hand,
generalizes previous quantum results for the case of zero field
\cite{Kuz91,haug-etal.92} and static field \cite{morawetz94}. Furthermore,
we derive results for the polarization and screening properties and
longitudinal field fluctuations in a strong field of arbitrary
time--dependence.

\section{Basic physical problems and definitions}\label{sec_bas}
We consider the time evolution of a dense charged particle system
under the influence of a strong time--dependent electromagnetic field and
inter--particle correlations.
\subsection{Free particle motion}
It is instructive to recall first the
motion of classical free charges in an external field.
From integrating Newton's equation,
$m_a d {\bf v}_a/dt = e_a {\bf E}(t)$, we obtain the velocity change of
a particle with charge $e_a$ and mass $m_a$ in the field ${\bf E}(t)$
during a time interval $[t',t]$,
\begin{eqnarray}
\Delta{\bf v}_a(t,t') = \frac{e_a}{m_a}
\int_{t'}^{t}d{\bar t}\,{\bf E}({\bar t}),
\label{va}
\end{eqnarray}
and the field induced displacement
\begin{eqnarray}
\Delta{\bf r}_a(t,t') =
\frac{e_a}{m_a}
\int_{t'}^{t}d{{\tilde t}}\int_{t'}^{{\tilde t}}d{\bar t}\,{\bf E}({\bar t})
=\frac{e_a}{m_a}
\int_{t'}^{t}d{{\tilde t}}\int_{{\tilde t}}^{t}d{\bar t}\,{\bf E}({\bar t}).
\label{ra}
\end{eqnarray}
In the equations above, we have dropped contributions from the
acceleration and velocity at the initial moment $t'$ since they are not
related to the field.
A further important quantity is the average kinetic energy
\begin{eqnarray}
\overline{E_a^{{\rm kin}}} = \frac{m_a}{2}\frac{1}{T}\int_{0}^{T}dt\, v_a^2(t),
\label{e-kin-av}
\end{eqnarray}
where, in case of a periodic field, $T$ is the oscillation period.
Below we will also need the change of the relative velocity
of a particle pair $a,b$ gained in the field $E(t)$
\begin{eqnarray}
\Delta{\bf v}_{ab}(t,t') \equiv \Delta{\bf v}_a(t,t') -
\Delta{\bf v}_b(t,t')
= \left( \frac{e_a}{m_a}-\frac{e_b}{m_b}\right)
\int_{t'}^{t}d{\bar t}\,{\bf E}({\bar t}),
\label{wab}
\end{eqnarray}
and the change of the inter--particle distance $\Delta{\bf r}_{ab}$
\begin{eqnarray}
\Delta{\bf r}_{ab}(t,t') \equiv \Delta{\bf r}_a(t,t') -
\Delta{\bf r}_b(t,t')
=\left( \frac{e_a}{m_a}-\frac{e_b}{m_b}\right)
\int_{t'}^{t}d{{\tilde t}}\int_{{\tilde t}}^{t}d{\bar t}\,{\bf E}({\bar t}).
\label{rab}
\end{eqnarray}

The most important special case is that of a
{\em harmonic time dependence}
\begin{equation}
{\bf E}(t) = {\bf E}_0 \cos \Omega t,
\label{efeld}
\end{equation}
which leads to the following explicit results for the quantities
introduced above
\begin{eqnarray}
\Delta{\bf v}_a(t,t') &=& {\bf v}_a^0
\left[\sin \Omega t - \sin \Omega t' \right],
\label{va-h}
\\
\Delta{\bf r}_a(t,t') &=& {\bf r}_a^0 \left[\Omega (t-t')\sin \Omega t
+\cos \Omega t - \cos \Omega t'\right].
\label{va-ra-h}
\end{eqnarray}
Here, we introduced two important quantities, the ``quiver'' velocity
\begin{eqnarray}
{\bf v}_a^0 \equiv \frac{e_a {\bf E}_0}{m_a \Omega},
\label{va0-h}
\end{eqnarray}
and the so--called excursion amplitude
\begin{eqnarray}
{\bf r}_a^0 \equiv \frac{e_a {\bf E}_0}{m_a \Omega^2}.
\label{ra0-h}
\end{eqnarray}
Furthermore, the cycle averaged ($T=2\pi/\Omega$) kinetic energy gain
of a charged particle (\ref{e-kin-av}) is the so--called ponderomotive
energy
\begin{eqnarray}
\overline{E_a^{{\rm kin}}} = \varepsilon^{\rm pond}_a =
\frac{e^2_aE_0^2}{4m_a \Omega^2}.
\label{e-pond-h}
\end{eqnarray}

Obviously, the above results may be extended to electromagnetic fields
with arbitrary time dependence, for example by expanding the field
in terms of harmonic components. Nevertheless, it is useful to explicitly
consider a second situation frequently encountered in modern applications:
pulsed fields, e.g. those produced by femtosecond lasers. We will
consider pulses of the following form
\begin{eqnarray}
{\bf E}(t) = {\bf E}_p(t)\cos \Omega t, \quad
{\bf E}_p(t) = 2 {\bf E}_0 \sin \Omega_p t, \quad 0\le \Omega_p t \le \pi,
\label{e-pulse}
\end{eqnarray}
and $E_p \equiv 0$ otherwise. Typically, $\Omega_p \ll \Omega$, although
modern femtosecond laser pulses may be as short as a few periods of the
main frequency $\Omega$. For the field (\ref{e-pulse}), we obtain the
velocity change and displacement
\begin{eqnarray*}
\Delta{\bf v}_a(t,t') &=& -\sum_{s=\pm}{\bf v}_a^{0s}
\left[\sin \Omega^s t - \sin \Omega^s t' \right], \quad
{\bf v}_a^{0\pm} \equiv \frac{e_a {\bf E}_0}{m_a \Omega^{\pm}}, \quad
\Omega^{\pm} \equiv \Omega_p \pm \Omega,
\\
\Delta{\bf r}_a(t,t') &=&  \sum_{s=\pm}{\bf r}_a^{0s}
\left[-\Omega^s (t-t')\cos \Omega^s t
+\sin \Omega^s t - \sin \Omega^s t'\right],
\end{eqnarray*}
with ${\bf r}_a^{0\pm} \equiv e_a {\bf E}_0/m_a \Omega^{\pm\,2}$,
whereas the change of relative velocity and two--particle distance,
$\Delta {\bf v}_{ab}$ and $\Delta {\bf r}_{ab}$ follow from
$\Delta {\bf v}_{a}$ and $\Delta {\bf r}_{a}$  by replacing
${\bf v}_a^{0\pm}$ by
${\bf v}_a^{0\pm}-{\bf v}_b^{0\pm}$ and ${\bf r}_a^{0\pm}$ by
${\bf r}_a^{0\pm}-{\bf r}_b^{0\pm}$, respectively.

The above results trivially include the case of a time-independent
electric field which is recovered by letting in Eq. (\ref{efeld})
$\Omega \rightarrow 0$. The corresponding results are
\begin{eqnarray*}
\Delta{\bf v}_a(t,t') = \frac{e_a}{m_a}{\bf E}_0 (t-t'), \qquad
\Delta{\bf r}_a(t,t') = \frac{1}{2}\frac{e_a}{m_a}{\bf E}_0 (t-t')^2.
\label{va-ra-s}
\end{eqnarray*}

It is instructive to consider a number of parameters which characterize
the state of the plasma, field strength, quantum properties etc:
\begin{enumerate}
\item{The field strength can be characterized by the ratio
$\alpha = v_a^0/v_{{\rm th},a}$ of the amplitude
of the oscillation velocity (quiver velocity) $v^0_{a}$, Eq.~(\ref{va-h}),
to the thermal velocity $v_{{\rm th},a}=(kT/m_a)^{1/2}$.}
\item{The relative importance of the field and of particle--particle
interaction is characterized
by $\beta = r_a^0/r_D$, where $r^0_a$ is the amplitude of the field induced
displacement (\ref{va-ra-h}), and $r_D$ is the Debye radius
(interaction range).}
\item{The frequency of the field has to be compared to the
eigenfrequencies of the plasma, most importrantly, the electron
Langmuir (plasma) frequency, $\gamma = \Omega/\omega_{pl}$, which reflects
competition between field frequency and plasma density effects.}
\item{The relevance of collisional processes depends on the ratio
$\delta = \nu/\Omega$, where $\nu$ is the total collision frequency of
electrons in the plasma.}
\item{The photon energy is characterized by its ratio to the thermal
energy, $\hbar \Omega/kT$.}
\end{enumerate}
Modern lasers easily produce strong fields which satisfy the inequalities
$\alpha \gg 1$ and $\beta \gg 1$. In high--frequency fields and/or
plasmas of moderate density, $\delta \ll 1$, which allows to treat
collisions perturbatively, see below.

\subsection{Two--particle scattering}
Coulomb interaction between the charged carriers as well as
quantum effects, obviously, may drastically modify the free particle
behavior. Scattering of two particles with charges $e_a, e_b$ in quantum
states $|{\bf k}_1\rangle$
and $|{\bf k}_2\rangle$ on the Coulomb potential
\begin{equation}
V_{ab}(q)=\frac{4\pi e_a e_b\hbar^2}{q^2},
\label{coulomb}
\end{equation}
leads to a
transfer of momentum ${\bf q}$ between them, so after the collision time
$t\sim t_{{\rm coll}}$ the particles are in momentum states
$|{\bf k}_1+{\bf q}\rangle$ and $|{\bf k}_2-{\bf q}\rangle$.
While conventional
kinetic approaches treat collision as instantaneous,
$t_{{\rm coll}}\rightarrow 0$, this is not appropriate for correlated
plasmas as well as in the presence of rapidly varying fields with
$\Omega \cdot t_{{\rm coll}}$ not being small. In this
case, during the collision time, the scattering partners will be
accelerated by the external field which is called {\em intra--collisional
field effect}, which essentially modifies the scattering process. Using
a quantum language, during the collision time, the particles may absorb
photons of the electromagnetic field which is the familiar
{\em inverse bremsstrahlung}, or re--emit them (bremsstrahlung).

The kinetic treatment of two--particle scattering on the Coulomb
potential (\ref{coulomb}) leads to the well--known divergencies at
short and long wavelengths. While the first is naturally cured by a
quantum theoretical approach, the origin of the latter is the long
range of the Coulomb interaction. The familiar solution lies in the
replacement of the bare Coulomb potential (\ref{coulomb}) by a screened one
\begin{eqnarray}
V_{ab}(q) \rightarrow V^s_{ab}(\omega,q)=
\frac{V_{ab}(q)}{\epsilon^R(\omega,q)},
\label{vs1}
\end{eqnarray}
where $\epsilon^R$ is the retarded dielectric function. This is not
only of fundamental interest but has also practical relevance.
The dielectric function includes collective plasma oscillations and
instabilities which, especially in nonequilibrium situations, may strongly
enhance scattering, transport and
energy exchange with the electromagnetic field (anomalous transport).

The simplest approximation for the dielectric function is the random phase
approximation (RPA) being the quantum generalization of the Vlasov dielectric
function
\begin{eqnarray}
\epsilon^R(\omega,q;t) &=& 1 - \sum_{a} V_{aa}(q)\, \Pi_a^R(\omega,q;t),
\label{df-rpa}
\\
\Pi^R_a(\omega,q;t) &=& \frac{1}{\hbar} \int \frac{d^3 k}{(2\pi\hbar)^3}
\frac{f_a({\bf k};t)-f_a({\bf k+q};t)}{\hbar \omega + i\delta
+\epsilon_a({\bf k})-\epsilon_a({\bf k+q})},
\end{eqnarray}
where $\epsilon_a$ denotes the single--particle energy.
Interestingly, this result was derived by Klimontovich and
Silin \cite{KS52a,KS52b} two years before Lindhard \cite{KS60}.
While this approximation is applicable to nonequilibrium situations
in which the Wigner distributions $f_a$ are weakly time--dependent
(when the time scale $t$ is much longer than $2\pi/\omega$),
on short times or for fast processes, such as in high frequency
fields, generalizations are necessary. Such generalizations
avoid the assumption of separation of the two time scales and
lead to an explicit dependence of the functions $\Pi, \epsilon$ and
$V^s$ on two times. Moreover, the presence of an electromagnetic field
may be expected to modify the dielectric and screening properties,
leading to a very complex problem of coupled particle, screening and
field dynamics. The appropriate theoretical concept to tackle this
problem is provided by quantum field theory.

\section{Quantum field theoretical approach to the dynamics of
plasmas in electromagnetic fields}
Numerous concepts have been developed to describe the mentioned
above dynamics of particles and fields. Among them, the most
systematic and powerful is the theory of nonequilibrium Green's
functions. It is based on the method of relativistic quantum field theory,
where charged particles and the longitudinal and transverse electromagnetic
field are described on
equal footing by field operators \cite{DuB67,BD72}. From the equations of
motion for the
field operators - the Dirac equation and Maxwell's equations, one
can derive equations of motion for all quantities of interest. Among
them, the most important are two--time correlation functions (Green's
functions) which allow for systematic and far--reaching generalizations
of traditional kinetic
theory. In this paper, we focus on nonrelativistic particle dynamics
and start our derivations from the familiar equations of motion
for the particle correlation functions $g^>$ and $g^<$ while
the electromagnetic field is treated classically.

\subsection{Kadanoff--Baym Equations}\label{kbe_ss}
The field theoretical description of plasmas is based on the
creation and annihilation operators $\psi^{\dagger}$ and $\psi$ \cite{n-klim}
which are defined to guarantee the spin statistics theorem,
\begin{eqnarray*}
\psi_{a}(1)\psi_{b}(2)\mp
\psi_{b}(2)\psi_{a}(1)
&=&
\psi^{\dagger}_{a}(1)\psi^{\dagger}_{b}(2)
\mp \psi^{\dagger}_{b}(2)\psi^{\dagger}_{a}(1)=0,
\nonumber\\
\psi_{a}(1)\psi^{\dagger}_{b}(2)
\mp \psi^{\dagger}_{b}(2)\psi_{a}(1)
&=& \delta(1-2)\,\delta_{a,b},
\label{psi-comm}
\end{eqnarray*}
where $t_1=t_2$ has been assumed. The upper (lower) sign refers to bosons
(fermions), $1\equiv ({\bf r}_1,t_1,s_1^{3})$, and $a$ labels the particle
species.
Below, we will drop the spin index and assume fermions.
The nonequilibrium state of a correlated plasma is described by the
two--time correlation functions which are statistical averages (with the
initial density operator of the system) of field operator products
\begin{eqnarray}
g_a^>(1,1')=\frac{1}{i\hbar}\langle\psi_a(1)\psi_a^{\dagger}(1')\rangle\,,
\qquad g_a^<(1,1')=
-\frac{1}{i\hbar}\langle\psi_a^{\dagger}(1')\psi_a(1)\rangle,
\label{ggtls_def}
\end{eqnarray}
where
$g_a^{>}$ and $g_a^{<}$ are, in nonequilibrium, independent from one another.
They contain the complete dynamical and statistical information. The latter
follows from their elements {\em along the time diagonal}: the one-particle
density matrix is immediately obtained from the function $g^<$
according to
\begin{equation}
f_a({\bf r}_1,{\bf r}_1',t) = -i\hbar g_a^{<}(1,1')|_{t_1=t'_1},
\label{f-def}
\end{equation}
whereas the dynamical information (e.g. the single--particle spectrum and
the correlations)
follows from the function values {\em across the diagonal} in the
$t_1-t_1'$--plane, in particular,
from the {\em spectral function} $a(1,1')$,
\begin{eqnarray}
a(1,1') \equiv i\hbar\left\{g_a^{>}(1,1')-g_a^{<}(1,1') \right\}
 =  i\hbar \left\{g_a^R(1,1')-g_a^A(1,1')\right\},
\label{a-def}
\end{eqnarray}
where $g^{R/A}$ are the retarded and advanced Green's functions,
defined below in Eq. (\ref{gra-def}).
In the following, it will often be convenient to use microscopic
and macroscopic time and space variables being defined as
\begin{eqnarray}
&&{\bf r}={\bf r}_1-{\bf r}'_1,\qquad {\bf R}=({\bf r}_1+{\bf r}'_1)/2,
\label{r-r-def}
\\
&&\tau=t_1-t'_1,\qquad t=(t_1+t'_1)/2\,.
\label{t-tau-def}
\end{eqnarray}
In particular, in cases where the microscopic variables vary on much smaller
scales than the
macroscopic ones, it is advantageous to perform a Fourier transformation
with respect to $\tau$ and/or ${\bf r}$ which leads to the frequency and
momentum variables $\omega$ and ${\bf p}$, respectively. In particular,
Eq. (\ref{f-def}) then yields the familiar Wigner distribution function
\begin{equation}
f_a({\bf p},{\bf R},t)=
-i\hbar g_a^{<}({\bf p},{\bf R};t_1,t'_1)\big|_{t_1=t'_1=t}.
\label{f-def2}
\end{equation}

The time evolution of the correlation functions in an electromagnetic
field is determined by the Kadanoff--Baym equations
\cite{kadanoff-baym,ksb85}
\begin{eqnarray}
&&\left[i\hbar\frac{\partial}{\partial t_1}
-\frac{1}{2m_a}\left(\frac{\hbar}{i}\nabla_1-\frac{e_a}{c}
{\bf A}(1)\right)^2 - e_a \phi(1)\right]g_a^{\gtrless}(1, 1')
- \int d{\bar {\bf r}}_1 \,
\Sigma_a^{\rm HF}(1,{\bar {\bf r}}_1t_1)
g_a^{\gtrless}({\bar{\bf r}}_1t_1,1')
\nonumber\\
&&=\int_{t_0}^{t_1} d{\bar 1} \,
\left[\Sigma_a^>(1,{\bar 1})-\Sigma_a^<(1,{\bar 1})\right] \,
g_a^{\gtrless}({\bar 1}, 1') -
\int_{t_0}^{t'_1} d{\bar 1} \,
\Sigma_a^{\gtrless} (1,{\bar 1})\,\left[g_a^>({\bar 1}, 1')-
g_a^<({\bar 1} ,1')\right],
\label{kb_eq}
\end{eqnarray}
which have to be fulfilled together with the adjoint equations.
Here, $t_0$ denotes the initial time where the system is assumed to be
uncorrelated
(otherwise, the equations have to be
supplemented with an initial correlation contribution to $\Sigma_a$, cf.
\cite{semkat-etal.99pre}). $\Sigma_a^{{\rm HF}}$ is the Hartree--Fock
selfenergy (mean--field energy with exchange),
\begin{eqnarray}
\Sigma_a^{{\rm HF}}(11') =
-i\hbar \delta(t_1-t_{1'})\sum_b \left\{
\int d {\bf r}_2 V_{ab}({\bf r}_1-{\bf r}_2) g_b^{<}(22^+)
- \delta_{a,b}V_{ab}({\bf r}_1-{\bf r}_1')g_b^{<}(11')\right\},
\label{sig_hf}
\end{eqnarray}
and $\Sigma_a^{\gtrless}$ are the correlation selfenergies
which will be discussed below.

For the following derivations, it is useful to introduce, in addition,
the retarded and advanced Green's functions
\begin{equation}
g_a^{R/A}(1,1')=
\pm \Theta[\pm(t_1-t'_1)]\left\{g_a^>(1,1')-g_a^<(1,1')\right\},
\label{gra-def}
\end{equation}
which obey the simpler equations
\begin{eqnarray}
&&\left[i\hbar\frac{\partial}{\partial t_1}-\frac{1}{2m_a}
\left(\frac{\hbar}{i}\nabla_1-\frac{e_a}{c}{\bf A}(1)\right)^2
- e_a \phi(1)\right]
g_a^{R/A}(1,1')
\nonumber\\
&&-\int d2 \,\Sigma_a^{R/A}(1,2)g_a^{R/A}(2,1')\,\, =\,\,
\delta(1-1').
\label{gra_eq}
\end{eqnarray}
In equations (\ref{kb_eq}) and (\ref{gra_eq}), the electromagnetic field is
given by the vector and scalar potentials ${\bf A}$ and $\phi$ and
will be treated classically. ${\bf A}$ is the full vector potential
(external plus induced) which obeys Maxwell's equations, whereas $\phi$
is understood as to be due to external sources only, the induced longitudinal
field is fully accounted for in the screened Coulomb potential $V^s$ which
enters the selfenergies $\Sigma_a^{>}$ and $\Sigma_a^{<}$,
see below.

Although one can directly analyze and solve the two--time
Kadanoff--Baym equations (\ref{kb_eq}), e.g. \cite{BKS+96},
for an overview, see \cite{bonitz-book}, it is easier to consider the
kinetic equation for the Wigner distribution function (\ref{f-def2}),
which we will be concerned with here.
This equation is immediately obtained from the equal time limit,
$t_1=t'_1=t$, of
Eq.~(\ref{kb_eq}) plus its adjoint. Introducing further the variables
${\bf R}$ and ${\bf r}$, Eqs.~(\ref{r-r-def}) and (\ref{t-tau-def}), we obtain, after Fourier
transformation with respect to ${\bf r}$, for the spatially homogeneous case,
\begin{eqnarray}
\frac{\partial}{\partial t}
f_a({\bf p},t)\, =\,- 2\, {\rm Re}\int_{t_0}^t d {\bar t}
\Big\{
\Sigma_a^{>}\left({\bf p};t,{\bar t}\right)\,
g_a^{<}\left({{\bf p};\bar t},t\right)-
\Sigma_a^{<}\left({\bf p};t,{\bar t}\right)\,
g_a^{>}\left({\bf p};{\bar t},t\right)\Big\}\,.
\label{f_eq}
\end{eqnarray}
This is an exact equation and, therefore, well suited for deriving
generalized kinetic equations.
However, this equation is not closed yet since it contains under the
collision integral functions depending on two times.
Therefore, to obtain explicit expressions for the collision integral,
one has:
\begin{enumerate}
\item to find appropriate approximations for the self
energy. For this, the Green's functions approach provides powerful
approximation schemes based on Feynman diagrams which allow for a
very systematic development of the theory. Here, we are interested in
the plasma dynamics with screening effects properly included. Therefore,
the appropriate choice for the selfenergy will be the random phase
approximation (RPA);
\item to express the correlation functions
$g_a^{\gtrless}$ as functionals of the Wigner functions $f_a$
(reconstruction problem). This problem can be solved approximately
on the basis of the generalized Kadanoff--Baym ansatz (GKBA) of
Lipavsk\'{y} et al.
\cite{lipavski-etal.86}, see Sec. \ref{gkba_ss}.
\end{enumerate}

\subsection{Gauge--invariant Green's functions}\label{gauge_ss}
It is well known that the electromagnetic field can be introduced in
various ways (gauges) which may lead to essentially different explicit
forms of the resulting kinetic equations.
Although alternative derivations are successfully applied too,
gauge invariance becomes a particular problem if the resulting
kinetic equations are treated by means of approximations,
such as retardation or gradient expansions. A critical issue is that
the result of these approximations maybe essentially different in different
gauges, see e.g. \cite{haug-jauho96} for examples. To
avoid these difficulties, we will formulate the theory in terms of
correlation functions which are made explicitly gauge--invariant. While
the main results have been presented in Ref. \cite{kremp-etal.99pre}, here we
provide some additional details.

In this section, we use a co-variant 4-vector notation as it makes
the following transformations more compact and symmetric. The
corresponding definitions are
\begin{eqnarray}
A_{\mu} = (c\phi,{\bf A}), \quad
x_{\mu} = ({c\tau,\bf r}), \quad
X_{\mu} = ({ct,\bf R}),
\nonumber
\end{eqnarray}
and the convention $a_{\mu}b^{\mu}=a_0b_0 - {\bf a}{\bf b}$ is
being used.

One readily proofs that the Kadanoff--Baym equations (\ref{kb_eq}) remain
covariant under gauge transformations, i.e., under the following
transformations of the potentials and field operators
\begin{eqnarray}
A'_\mu(x)=A_\mu(x)-\partial_\mu\chi(x)\qquad
\psi_a'(x)=e^{\frac{i}{\hbar}\frac{e_a}{c}\chi(x)} \psi_a(x),
\label{gauge-a}
\end{eqnarray}
The corresponding gauge transform of the Green's functions leads to
\begin{eqnarray*}
g_a'(x,X)=e^{\frac{i}{\hbar} \frac{e_a}{c}\left[\chi\left(X+\frac{x}{2}\right)
-\chi\left(X-\frac{x}{2}\right)\right]} g_a(x,X)\,.
\end{eqnarray*}
Following an idea of Fujita \cite{fujita66}, we now introduce a gauge--invariant
Green's function $g({\bf k},X)$ which is given by the modified Fourier transform
\begin{eqnarray}
g_a({\bf k},X)=\int
\frac{d^4x}{(2\pi)^4}\exp\left\{i\int_{-\frac{1}{2}}^{\frac{1}{2}}
d\lambda \, x_\mu \left[k^\mu+\frac{e_a}{c}A^\mu(X+\lambda x)\right]\right\}
g_a(x,X),
\label{gauge-ft}
\end{eqnarray}
where use has been made of the identity
\[
\chi\left(X+\frac{x}{2}\right) -\chi\left(X-\frac{x}{2}\right)
=\int_{-\frac{1}{2}}^{\frac{1}{2}} d\lambda \,
\frac{d}{d\lambda}\chi\left(X+\lambda x\right) =
x_{\mu} \partial^{\mu}\int_{-\frac{1}{2}}^{\frac{1}{2}} d\lambda \,
\chi\left(X+\lambda x\right).
\]
Indeed, one readily confirms that under any gauge transform (\ref{gauge-a}),
the phase
factors cancel, and $g'({\bf k},X) \equiv g({\bf k},X)$, \cite{haug-jauho96}.

In the following, we focus on spatially homogeneous electric fields and use
the vector potential gauge
\begin{eqnarray}
A_0=\phi=0;\qquad {\bf A}= - c \int^t_{-\infty} d{\bar t}\,{\bf E}({\bar t}).
\label{v-gauge}
\end{eqnarray}
In this case, relation (\ref{gauge-ft}) simplifies to

\begin{eqnarray}
g_a({\bf k},\omega;{\bf R},t)=\int d\tau d{\bf r}\exp \left[i\,\omega \tau -
\frac{i}{\hbar}{\bf r}\cdot \left({\bf k}+
\frac{e_a}{c}\int\limits_{t-\frac{\tau}{2}}^{t+\frac{\tau}{2}}
\frac{dt'}{\tau}\,
{\bf A}(t')\right) \right] g_a({\bf r},\tau;{\bf R},t),
\label{gauge-ftv}
\end{eqnarray}
what means that the gauge--invariant Green's function $g({\bf k})$ follows
from the Wigner transformed function $g_a ({\bf p})$  by replacing the
canonical momentum ${\bf p}$ by the gauge--invariant kinematic momentum
${\bf k}$ according to
\begin{eqnarray}
{\bf p}={\bf k}+
\frac{e_a}{c}\int\limits_{t-\frac{\tau}{2}}^{t+\frac{\tau}{2}} \,
dt' \,\frac{{\bf A}(t')}{\tau}.
\label{kin-mom}
\end{eqnarray}
Let us illustrate this for the examples studied in Sec.~\ref{sec_bas}.
For a {\em harmonic electric field} given by Eq.~(\ref{efeld}),
the vector potential and the momentum relation become,
according to Eq.~(\ref{v-gauge}),
\begin{eqnarray}
{\bf A}(t) = -\frac{c {\bf E}_0}{\Omega}\sin \Omega t, \qquad
\label{a-h}
{\bf p} ={\bf k}+\frac{2}{\tau} \frac{{\bf E}_0}{\Omega^2} \,\sin
\Omega t
\,\sin \frac{\Omega \tau}{2}.
\end{eqnarray}
Similarly, for a {\em pulsed field} of the form Eq.~(\ref{e-pulse}),
the result for the vector potential and the momenta is
\begin{eqnarray*}
{\bf A}(t) = c {\bf E}_0\sum_{s=\pm}
\frac{\cos \Omega^s t}{\Omega^s}, \qquad
{\bf p} = {\bf k}+\frac{2{\bf E}_0}{\tau} \,
\sum_{s=\pm}\frac{1}{\Omega^{s\,2}}\cos\Omega^s t\,\sin \frac{\Omega^s \tau}{2}.
\end{eqnarray*}
Finally, for a {\em static field} we obtain,
\begin{eqnarray*}
{\bf A}(t) = -c {\bf E}_0 t, \qquad
{\bf p} ={\bf k} + e_a {\bf E}_0 t.
\end{eqnarray*}

For the derivations below, we will need the gauge invariant Fourier
transform of the convolution of two functions which, in the homogeneous
case, is given by
\begin{eqnarray}
I({\bf r}_1-{\bf r}'_1;\,t_1,t'_1)=\int d\bar{t} d\bar{{\bf r}} \,
B({\bf r}_1-\bar{{\bf r}};\,t_1,\bar{t})\,
C(\bar{{\bf r}}-{\bf r}'_1;\,\bar{t},t'_1).
\label{conv}
\end{eqnarray}
After straightforward manipulations which involve the
back transform of (\ref{gauge-ftv}), we arrive at
\begin{eqnarray}
I({\bf k};\,t_1,t'_1)&=&\int d\bar{t}\,
B\left[{\bf k}+\frac{e_a}{c}\int_{t'_1}^{t_1} dt''\, \frac{{\bf A}
(t'')}{t_1-t'_1} -
\frac{e_a}{c}\int_{\bar t}^{t_1} dt'' \,
\frac{{\bf A}(t'')}{t_1-{\bar t}};\, t_1,{\bar t}\right]
\nonumber
\\
&&\qquad
\times\,\, C \left[{\bf k}+
\frac{e_a}{c}\int_{t'_1}^{t_1} dt''\, \frac{{\bf A}
(t'')}{t_1-t'_1} - \frac{e_a}{c}
\int^{\bar{t}}_{t'_1} dt''\, \frac{{\bf A}(t'')}
{\bar{t}-t'_1};\,\bar{t},t'_1\right].
\quad
\label{conv-ft}
\end{eqnarray}
In particular, the derivation of the collision integral in the kinetic
equation for the Wigner function, requires the equal--time limit of this
expression, $t_1=t'_1=t$,
\begin{eqnarray}
I({\bf k};\,t)&=&\int d\bar{t}\,
B\left[{\bf k}+\frac{e_a}{c}{\bf A}(t) -
\frac{e_a}{c}\int_{\bar t}^{t} dt'' \,
\frac{{\bf A}(t'')}{t-{\bar t}};\, t,{\bar t}\right]
\nonumber
\\
&&\qquad
\times\,\, C \left[{\bf k}+\frac{e_a}{c}{\bf A}(t) -
\frac{e_a}{c}\int_{\bar t}^{t} dt'' \,
\frac{{\bf A}(t'')}{t-{\bar t}}
;\,\bar{t},t\right],
\quad
\label{conv-ftt}
\end{eqnarray}
where $B$ and $C$ will be replaced by $g^{\gtrless}$ and $\Sigma^{\gtrless}$.
Notice that in this case
the momentum arguments of $B$ and $C$ are equal. To simplify the notation
below, we introduce the field induced momentum shift
\begin{equation}
{\bf K}^A_a(t,t') \equiv \frac{e_a}{c}\int_{t'}^{t} dt'' \,
\frac{{\bf A}(t)-{\bf A}(t'')}{t-{t'}},
\label{ka}
\end{equation}
which has the important property
\begin{equation}
{\bf K}^A_a(t,t') - {\bf K}^A_a(t',t) = \frac{e_a}{c}
\left\{ {\bf A}(t)-{\bf A}(t')\right\} =
- e_a\int_{t'}^{t} dt'' \, {\bf E}(t'') \equiv {\bf Q}_a(t,t'),
\label{ka-dif}
\end{equation}
where  ${\bf Q}_a$ is nothing but minus the momentum gain of a
free particle in the field,
${\bf Q}_a(t,t')= - m_a \Delta {\bf v}_a(t,t')$, cf. Eq.~(\ref{va}).
Another important relation follows from multiplication by the time interval:
\begin{equation}
\frac{1}{m_a}{\bf K}^A_a(t,t')\cdot(t-t')
 \equiv - {\bf R}_a(t,t'),
\label{ka-dt}
\end{equation}
where ${\bf R}_a$ is just the field induced displacement of a free particle,
Eq.~(\ref{ra}), i.e. ${\bf R}_a(t,t')=\Delta{{\bf r}_a(t,t')}$.

Definition (\ref{ka}) allows us to rewrite Eq.~(\ref{conv-ftt}) as
\begin{eqnarray}
I({\bf k};\,t)&=&
\int d\bar{t}\,B\left[{\bf k}+{\bf K}^A_a(t,{\bar t});\, t,{\bar t}\right]
\, C \left[{\bf k}+{\bf K}^A_a(t,{\bar t});\, t,{\bar t}\right].
\label{conv-ftt2}
\end{eqnarray}

\subsection{Gauge invariant propagator. Generalized Kadanoff--Baym ansatz}
\label{gkba_ss}
As noted in Sec.~\ref{kbe_ss}, for the derivation of the collision integral
in the kinetic equation, we need to express the functions $g^>$ and $g^<$
in terms of the Wigner function. In addition, such a reconstruction ansatz
involves the retarded and advanced Green's functions $g^{R/A}(t,t')$ for which
suitable expressions have to be found. We first determine these quantities for
free particles in an electromagnetic field which allows to simplify
Eq.~(\ref{gra_eq}) to
\[
\left[i\hbar\frac{\partial}{\partial t_1} - \frac{1}{2 m_a} \left\{
{\bf p}-\frac{e_a}{c}{\bf A}(t_1)\right\}^2\right]
g_{a}^{R/A}({\bf p};t_1,t'_1)=\delta (t_1-t'_1),
\]
which is solved immediately by
\begin{eqnarray}
g_{a}^{R}({\bf p};\tau,t)
=-\frac{i}{\hbar}\Theta(\tau)\exp
\left[-\frac{i}{\hbar}\int\limits_{t-\frac{\tau}{2}}^{t+\frac{\tau}{2}}
dt'
\, [{\bf p}-\frac{e_a}{c}{\bf A}(t')]^2/2m_a\right],
\label{gra_free}
\end{eqnarray}
and $g_{a}^{A}$ is obtained from the symmetry relation
$g_{a}^{A}({\bf p};\tau,t)=[g_{a}^{R}({\bf p};-\tau,t)]^*$.
From this result, we can calculate the spectral function $a(t,t')$,
Eq.~(\ref{a-def}),
\begin{eqnarray}
a_{a}({\bf p};\tau,t)=\exp
\left[-\frac{i}{\hbar}\int\limits_{t-\frac{\tau}{2}}^{t+\frac{\tau}{2}}
dt'\, [{\bf p}-\frac{e_a}{c}{\bf A}(t')]^2/2m_a\right].
\label{a_free}
\end{eqnarray}
Obviously, the results (\ref{gra_free}) and (\ref{a_free}) are gauge-dependent
since $g_{a}^{R/A}$ and $a_{a}$ are functions of the
canonic momentum ${\bf p}$. But one can easily obtain the corresponding
gauge--invariant results by applying
the transform (\ref{gauge-ft}), with the result
\begin{eqnarray}
g_{a}^{R}({\bf k};\tau,t) &=& - \frac{i}{\hbar}\Theta(\tau)\,
e^{-\frac{i}{\hbar}\left[\frac{k^2}{2 m_a}\tau + S_a({\bf A};\tau,t)\right]
},
\label{gra_fg}
\\
\nonumber\\
\mbox{where}\quad S_a({\bf A};\tau,t) &=&
\frac{e_a^2}{2 m_a c^2} \left[
\int_{t-\frac{\tau}{2}}^{t+\frac{\tau}{2}} dt' A^2(t')
+ \frac{1}{\tau}
\left(\int_{t-\frac{\tau}{2}}^{t+\frac{\tau}{2}} dt'
{\bf A}(t')\right)^2\right]\,.
\end{eqnarray}
This result has a simple physical interpretation. For a free particle
without field, the spectral function shows free undamped oscillations
along $\tau$ (i.e. perpendicular to the time diagonal)
with the one--particle energy $\epsilon_a(k)=k^2/2m_a$, and its Fourier
transform is
\begin{equation}
a^{{\rm free}}_a({\bf k};\omega,t)=\delta[\hbar\omega-\epsilon_a(k)].
\label{afree}
\end{equation}
This clearly underlines the meaning
of the functions $g^{R/A}$ and, in particular, the spectral function $a$ --
they contain the full
information on the single--particle energy spectrum. Furthermore, in
a correlated system, the single--particle spectrum is affected by interactions
with other particles. This leads to a shift of the oscillation frequency
and to damping of the oscillations, i.e. to
finite life time effects, and it is reasonable to call the corresponding
single--particle excitations {\em quasi--particles}. On the other hand, the
result (\ref{gra_fg}) reflects the influence of an electromagnetic field on the
particle spectrum, while correlation effects have been neglected.
Eq.~(\ref{gra_fg}) shows that the field causes a time--dependent shift of
the single--particle energy which, obviously, reflects the well--known
fact that the proper eigenstates of the system contain the electromagnetic
field and are given by Volkov states \cite{Vol35}.
The spectrum may even contain additional peaks which becomes particularly
transparent in the limiting case of a harmonic time dependence:
For the field (\ref{efeld}),
the time integrations in $S$ can be performed, and
simple trigonometric relations lead to \cite{johnson-etal.96}
\begin{eqnarray}
S_a({\bf A};\tau,t)=
\varepsilon_a^{\rm pond}\,\tau\left[1 -
\frac{\sin
\Omega\tau\cos 2\Omega t}{\Omega \tau} + \frac{8\sin^2\Omega t \sin^2
\frac{\Omega \tau}{2}} {(\Omega\tau)^2}\right],
\label{gra_fgh}
\end{eqnarray}
where $\varepsilon_a^{\rm pond}$ is the ponderomotive
potential which was introduced in Eq.~(\ref{e-pond-h}). The first term
in the brackets leads to a shift of the single-particle energy, the average
kinetic energy of the particles increased by $\varepsilon_a^{\rm pond}$.
The remaining terms modify the spectrum qualitatively giving rise to
additional peaks which are related to {\em photon sidebands}
\cite{johnson-etal.96}.

Now we turn to the solution of the reconstruction problem. The simplest
solution is the common {\em Kadanoff--Baym ansatz},
\begin{eqnarray}
\pm i\hbar g_a^{\gtrless}({\bf p};\omega,t)=
a^{{\rm free}}_a({\bf p};\omega,t)f^{\gtrless}({\bf p};t),
\label{kba}
\end{eqnarray}
where $f^<\equiv f$ and $f^> \equiv 1 - f$, and the upper (lower) sign refers
to $g^>$ ($g^<$). Indeed, the two--time functions
$g_a^{>}(t_1,t'_1)$ and $g_a^{<}(t_1,t'_1)$ are now expressed in
terms of one--time Wigner distribution functions and a known spectral
function.
However, due to the expected retardation effects, this
ansatz is not applicable here.
As mentioned above, a more general solution which properly takes into
account retardation (memory) effects is the
{\em generalized Kadanoff--Baym ansatz} proposed by Lipavsk\'{y} et al.
\cite{lipavski-etal.86} which reads
\begin{eqnarray}
g_a^{\gtrless}({\bf p};t_1,t'_1)=
i\hbar g_a^R({\bf p};t_1,t'_1)g_a^{\gtrless}({\bf p};t'_1,t'_1)-
i\hbar g_a^{\gtrless}({\bf p};t_1,t_1)g_a^A({\bf p};t_1,t'_1),
\label{gkba}
\end{eqnarray}
where for the functions on the time diagonal
$\pm i\hbar g_a^{\gtrless}({\bf p};t_1,t_1)=f^{\gtrless}({\bf p};t_1)$,
cf. Eq.~(\ref{f-def2}).
Within the quasiparticle approximation and with static selfenergies,
Eq.~(\ref{gkba}) is exact. In more complex situations, it is
an approximation to the exact reconstruction solution, which has prooved
extremely successful in many applications. In particular, it has been used
for more general selfenergies and also with more general propagators
$g_a^{R/A}$, e.g. \cite{haug-jauho96}. We, therefore will use this ansatz
below.

Eq.~(\ref{gkba}) is written in terms of the momentum ${\bf p}$ and is,
therefore, gauge--dependent.
To transform this relation into a gauge--invariant form, we use
its coordinate representation,
\begin{eqnarray}
\pm g_a^{\gtrless}({\bf r}_1-{\bf r}'_1;t_1,t'_1) &=& \int d\bar{{\bf r}}\,
g_a^R({\bf r}_1-\bar{{\bf r}};t_1,t'_1) \,
f_a^{\gtrless}(\bar{{\bf r}}-{\bf r}'_1;t'_1)
\nonumber\\
&-& \int d\bar{{\bf r}}\,
f_a^{\gtrless}({\bf r}_1-\bar{{\bf r}};t_1) \,
g_a^A(\bar{{\bf r}}-{\bf r}'_1;t_1,t'_1),
\label{gkba_r}
\end{eqnarray}
and apply the gauge--invariant Fourier transform (\ref{gauge-ftv})
together with the back transforms of $g_a^{R/A}$ and $f_a^{\gtrless}$
which leads to the {\em gauge--invariant generalization of the GKBA}
\begin{eqnarray}
\pm g_a^{\gtrless}({\bf k};t_1,t'_1)=  g_a^R({\bf k};t_1,t'_1) \,
f_a^{\gtrless}\left[{\bf k}-{\bf K}^A_a(t',t); t'_1\right]
- f_a^{\gtrless}\left[{\bf k}-{\bf K}^A_a(t,t'); t_1\right] \,
g_a^A({\bf k};t_1,t'_1),
\label{gkba_gi}
\end{eqnarray}
where the definition (\ref{ka}) for ${\bf K}^A_a$ has been used.
As in the field--free case, the first term is nonzero only for $t_1 \ge t_1'$
and the second in the opposite case. Notice the difference of the time
arguments in the two distribution functions.

\section{General kinetic equation for quantum particles including screening
and electromagnetic fields}

We now come back to the time--diagonal limit of the  Kadanoff--Baym
equations, cf. Eqs.~(\ref{kb_eq}) and (\ref{f_eq}), and derive the
quantum kinetic equation for a plasma in a laser field thereby fully
taking into account dynamical screening. Again, it is
advantageous to derive this equation for the gauge--invariant
Wigner distribution. To this end, we take the Fourier transform
(\ref{gauge-ftv}) of the time--diagonal Kadanoff--Baym equation
\begin{eqnarray}
&&\frac{\partial}{\partial t} f_a({\bf k}_a,t)+e_a{\bf E}(t)\cdot
\nabla_{\bf k} f_a({\bf k}_a,t)=
 -2 {\rm Re}\int_{t_0}^t d{\bar t}\Big\{\Sigma_a^>g_a^< -
\Sigma_a^<g_a^>\Big\} = I_a({\bf k}_a,t), \qquad
\label{f_eqgi}
\end{eqnarray}
where the full arguments are, according to the
convolution relation
(\ref{conv-ftt2}), given by
\begin{eqnarray}\label{sig-g}
\Sigma_a^{\gtrless}g_a^{\lessgtr}  \equiv
\Sigma_a^{\gtrless}\left[{\bf k}_a+{\bf K}_a^A(t,{\bar t});t,{\bar t}\right] \,
g_a^{\lessgtr}\left[{\bf k}_a+{\bf K}_a^A(t,{\bar t});{\bar t},t\right].
\nonumber
\end{eqnarray}
This expression is valid for arbitrary approximations for the selfenergies
$\Sigma^>$ and $\Sigma^<$. In our previous paper \cite{kremp-etal.99pre},
we used the simple static Born approximation. Here, we are interested in a
fully selfconsistent {\em inclusion of dynamical screening}, so the appropriate
choice is the random phase approximation (RPA).

\subsection{Random phase approximation}\label{rpa_ss}
Starting from the familiar expression in
coordinate representation, application of the gauge--invariant Fourier
transform (\ref{gauge-ftv}) straightforwardly leads to
the following gauge--invariant result
\begin{eqnarray}
\Sigma_{a}^{\gtrless}({\bf k};t,t') =
i\hbar \int \frac{d {\bf q}} {(2\pi\hbar)^3}\,
g_a^{\gtrless}({\bf  k-q};t,t')\,
V_{aa}^{s\,\gtrless}({\bf q};t,t'),
\label{sigma-rpa}
\end{eqnarray}
which transforms the collision integral of Eq.(\ref{f_eqgi})
into
\begin{eqnarray}\label{coli1}
I_{a}({\bf k}_a,t) &=& - 2 \,{\rm Re}
\int \frac{d{\bf q}} {(2\pi\hbar)^3}\,
\int_{t_0}^t d {\bar t} \;\bigg\{
g_a^>\left[{\bf k}_a-{\bf q}+{\bf K}_a^A(t,{\bar t});t,{\bar t}\right]
\nonumber\\
&\times&
i\hbar V_{aa}^{s\,>}({\bf q};t,{\bar t})\,
g_a^<\left[{\bf k}_a+{\bf K}_a^A(t,{\bar t});{\bar t},t\right]\,
\quad - \quad [\;>\; \longleftrightarrow \;<\;] \bigg\}
\end{eqnarray}

In Eq.~(\ref{sigma-rpa}) we introduced the correlation functions of the
screened potential (plasmon correlation functions)
$V_{s}^{>}, V_{s}^{<},$
which contain the whole screening problem and are directly related to
the correlation function of the {\em longitudinal field fluctuations}
(microfield fluctuations) \cite{SB87}
\begin{eqnarray}
\frac{e_a^2 \hbar^2}{q^2}\;\overline{
\langle \delta E(t) \delta E(t') \rangle }_{{\bf q}} =
\frac{1}{2}
\left[i\hbar V_{aa}^{s\,>}({\bf q};t,t') +
i\hbar V_{aa}^{s\,<}({\bf q};t,t')\right].
\label{dede}
\end{eqnarray}
While in the classical case, the contributions from $V^>$ and $V^<$ are
equal, in the quantum case a symmetrization is useful which is indicated
by the bar over the fluctuation term.
$V^{s\,>}$ and $V^{s\,<}$ can be related to the retarded and advanced
screened potentials via the optical theorem
\begin{eqnarray}
V_{ab}^{s\,\gtrless}({\bf q};t_1,t_2) =
\sum_{c}\int_{t_0}^{t_1} dt_3 \int_{t_0}^{t_2} dt_4
V_{ac}^{s\,R}({\bf q};t_1,t_3)\,\Pi_{cc}^{\gtrless}({\bf q};t_3,t_4)
V_{cb}^{s\,A}({\bf q};t_4,t_2),
\label{opt}
\end{eqnarray}
where $V_s^{R}$ and $V_s^{A}$
obey the following equation of motion
(Dyson equation)
\begin{eqnarray}
V_{ab}^{s\,R/A}({\bf q};t,t') = V_{ab}({\bf q})\delta(t-t') + \sum_{c}
V_{ac}({\bf q})\int_{t'}^{t}d{\bar t} \,
\Pi_{cc}^{R/A}({\bf q};t,{\bar t}) V_{cb}^{s\,R/A}({\bf q};{\bar t},t'),
\label{vdyson}
\end{eqnarray}
and $V^{s\,R/A}$ are related to the nonequilibrium inverse dielectric
function according to
\begin{eqnarray}
V_{ab}^{s\,R/A}({\bf q};t,t') =
V_{ab}({\bf q})\left[\epsilon^{R/A}({\bf q};t,t')\right]^{-1}.
\label{idf}
\end{eqnarray}
In the above equations, $V_{ab}({\bf q})$ is the bare Coulomb potential
(\ref{coulomb}) and
$\Pi^{R/A}$ the retarded and advanced longitudinal polarization functions
(plasmon selfenergies).
To close this system of equations, the polarization functions have to
be expressed in terms of the particle correlation functions for which the
simplest approximation is provided by the RPA,
\begin{eqnarray}
\Pi^{\gtrless}_{bb}({\bf q};t_1,t_2) &=&
- i\hbar \int \frac{d^3 {\bf k}_b}{(2\pi\hbar)^3}\,
g^{\gtrless}_b({\bf k}_b+{\bf q};t_1,t_2)\,g^{\lessgtr}_b({\bf k}_b;t_2,t_1),
\label{pigl}\\
\Pi^{R}_{bb}({\bf q};t,t') &=& \Theta(t-t')
\left\{\Pi_{bb}^{>}({\bf q};t,t') - \Pi_{bb}^{<}({\bf q};t,t')
\right\},
\label{pira}
\end{eqnarray}
and $\Pi^A$ follows from the relation
$\Pi^{A}_{bb}({\bf q};t,t')=[\Pi^{R}_{bb}({\bf q};t',t)]^*$.
This set of equations completely defines the non--Markovian
polarization approximation (RPA) for a quantum plasma in a strong
transverse field.

\subsection{Application of the gauge--invariant GKBA}
What is left to obtain a closed expression for the
collision integral in equation (\ref{f_eqgi}) is to apply the
gauge--invariant GKBA (\ref{gkba_gi}) together with
the free--particle approximation (\ref{gra_fg}) to all two--time functions.
This leads to the following results for the optical
theorem:
\begin{eqnarray}
V_{aa}^{s\,\gtrless}({\bf q};t_1,t_2) &=&
-\frac{i}{\hbar}\sum_b\int\frac{d^3k}{(2\pi\hbar^3)}
\int_{t_0}^{t_1} dt_3 \int_{t_0}^{t_2} dt_4
V_{ab}^{s\,R}({\bf q};t_1,t_3)\,V_{ba}^{s\,A}({\bf q};t_4,t_2)\,
\nonumber\\
&\times&\bigg\{
\Theta(t_3-t_4)\,e^{-\frac{i}{\hbar}\left[
\left(\epsilon^b_{{\bf k}+{\bf q}}-\epsilon^b_{{\bf k}}\right)(t_3-t_4)
-{\bf q}{\bf R}_b(t_3,t_4)
\right]}
\nonumber\\
&&\times
f_b^{\gtrless}\left[{\bf k}+{\bf q}+{\bf Q}_b(t_3,t_4);t_4\right]\,
f_b^{\lessgtr}\left[{\bf k}+{\bf Q}_b(t_3,t_4);t_4\right]
\nonumber\\
&&+
\Theta(t_4-t_3)\,e^{-\frac{i}{\hbar}\left[
\left(\epsilon^b_{{\bf k}+{\bf q}}-\epsilon^b_{{\bf k}}\right)(t_3-t_4)
+{\bf q}{\bf R}_b(t_4,t_3)
\right]}
\nonumber\\
&&\times
f_b^{\gtrless}\left[{\bf k}+{\bf q}+{\bf Q}_b(t_4,t_3);t_3\right]\,
f_b^{\lessgtr}\left[{\bf k}+{\bf Q}_b(t_4,t_3);t_3\right]
\bigg\}.
\label{vs-gkba}
\end{eqnarray}
Similarly, the result for $\Pi^{R}$ can be transformed to
\begin{eqnarray}
\Pi^{R}_{aa}({\bf q};t,t') &=&
-\frac{i}{\hbar}\Theta(t-t')\,
e^{\frac{i}{\hbar} {\bf q}{\bf R}_a(t,t')}
\int \frac{d^3 k}{(2\pi\hbar)^3}\,
e^{-\frac{i}{\hbar}
\left(\epsilon^a_{{\bf k+q}}-\epsilon^a_{{\bf k}}\right)(t-t')
}
\nonumber\\
&\times &
\left\{ f_a\left[{\bf k}+{\bf Q}_a(t,t');t'\right]-
f_a\left[{\bf k+q}+{\bf Q}_a(t,t');t'\right]
\right\},
\label{pira-gkba}
\end{eqnarray}
where the momentum shift ${\bf Q}_a$ and field--induced displacement
${\bf R}_a$
were defined above in Eqs.~(\ref{ka-dif}) and Eqs.~(\ref{ka-dt}),
respectively.
In the absence of the electromagnetic field, (${\bf Q}_a \rightarrow 0,
{\bf R}_a \rightarrow 0$), Eq.~(\ref{pira-gkba}) reduces
to the well--known nonequilibrium RPA--polarization function. The effect
of the field is two--fold: first, it introduces an additional retardation
${\bf Q}_a$ in the distributions (intra--collisional field effect) and second,
it leads to a modification of the one--particle energies in the exponent
given by ${\bf R}_a$ which we discussed in detail in ref.
\cite{kremp-etal.99pre}.

We now can transform the collision integral
Eq.~(\ref{coli1}) by applying the GKBA (\ref{gkba_gi})
to $g^>, g^<$,
and using the result for $V^{s>}$ and $V^{s<}$, Eq.~(\ref{vs-gkba}). After
straightforward calculations, we obtain the final result
\begin{eqnarray}\label{coli3}
I_{a}({\bf k}_a,t) &=& - \frac{2}{\hbar^2} \sum_b\,{\rm Re}
\int \frac{d{\bf k}_b d{\bar {\bf k}}_a d{\bar {\bf k}}_b}{(2\pi\hbar)^6}
\,\delta ({\bf k}_{a}+{\bf k}_{b}-\bar{{\bf k}}_{a}-\bar{{\bf k}}_{b})\,
\int_{t_0}^t d {\bar t}
\nonumber\\
&\times& \Bigg\{
e^{-\frac{i}{\hbar}\left[
\left({\bar \epsilon}_a-\epsilon_a\right)(t-{\bar t})
+({\bf k}_a-{\bar {\bf k}}_a){\bf R}_a(t,{\bar t})
\right]}\,
f_a^>\left[{\bar {\bf k}}_a+{\bf Q}_a(t,{\bar t});{\bar t}\right]
f_a^<\left[{\bf k}_a+{\bf Q}_a(t,{\bar t});{\bar t}\right]\,
\nonumber\\
&\times&\int_{t_0}^t d t_3 \;\int_{t_0}^{\bar t} d t_4
V^{s\,R}({\bf k}_a-{\bar {\bf k}}_a;t_1,t_3)\,
V^{s\,A}({\bf k}_a-{\bar {\bf k}}_a;t_4,t_2)\,\times
\nonumber\\
&&\bigg[
\Theta(t_3-t_4)\,e^{-\frac{i}{\hbar}\left[
\left({\bar \epsilon}_b-\epsilon_b\right)(t_3-t_4)
+({\bf k}_b-{\bar {\bf k}}_b){\bf R}_b(t_3,t_4)
\right]}
\nonumber\\
&&\times
f_b^>\left[{\bar {\bf k}}_b+{\bf Q}_b(t_3,t_4);t_4\right]\,
f_b^<\left[{\bf k}_b+{\bf Q}_b(t_3,t_4);t_4\right]
\nonumber\\
&&+
\Theta(t_4-t_3)\,e^{-\frac{i}{\hbar}\left[
\left({\bar \epsilon}_b-\epsilon_b\right)(t_3-t_4)
-({\bf k}_b-{\bar {\bf k}}_b){\bf R}_b(t_4,t_3)
\right]}
\nonumber\\
&&\times
f_b^>\left[{\bar {\bf k}}_b+{\bf Q}_b(t_4,t_3);t_3\right]\,
f_b^<\left[{\bf k}_b+{\bf Q}_b(t_4,t_3);t_3\right] \bigg]
\quad - \quad [\;>\; \longleftrightarrow \;<\;] \Bigg\}.
\end{eqnarray}
This is the general non--Markovian collision integral for a homogeneous
weakly coupled {\em dynamically screened plasma in an electromagnetic field}.
It is a generalization of numerous results which were previously obtained
by various authors, including
our static screening result for strong time--dependent fields
\cite{kremp-etal.99pre} and the result of Silin and Uryupin
\cite{silin-ur}, the RPA result
for a static field of Morawetz \cite{morawetz94} and
the field-free RPA results of Kuznetsov \cite{Kuz91} and Haug and
Ell \cite{haug-etal.92}. Furthermore, it generalizes previous results
obtained for classical plasmas by Silin, Oberman et al., Klimontovich
and others \cite{silin60,oberman-etal.62,klimontovich75}. In particular,
as we will see below, the classical dynamical screening result of
Klimontovich and Puchkov \cite{KP74} is straightforwardly recovered
from the collision integral (\ref{coli3}). This collision integral is
the basis for computing electron--ion
collision frequencies, plasma heating inverse bremsstrahlung effects etc.,
thereby fully taking into account dynamical screening, plasma instabilities
and anomalous transport

Despite the complicated structure of the integral (\ref{coli3}), a
direct numerical integration of the kinetic equation (\ref{f_eqgi})
appears to be within reach, as recently solutions of non--Markovian
RPA-type equations (without longitudinal field) for semiconductors have
been reported \cite{banyai-etal.98,kwong-etal.rpa,moldzio-rpa}.
On the other hand,
to gain deeper insight in the physical processes contained in the
collision term (\ref{coli3}), it is useful to consider analytical
simplifications.

\subsection{High frequency fields. Silin ansatz}\label{silin_ss}
If the collision frequency is low compared to the oscillation frequency
of the field, i.e. if the parameter $\delta \ll 1$, one may follow an
idea of Silin \cite{silin64} and solve the kinetic equation
(\ref{f_eqgi}) with a perturbation ansatz $f_a=f_a^0+f_a^1$ where
$f_a^0$ obeys the collisionless equation
\begin{eqnarray}
\frac{\partial}{\partial t} f^0_a({\bf k},t)+e_a{\bf E}(t)\cdot
\nabla_{\bf k} f^0_a({\bf k},t) = 0
\nonumber\\
\mbox{with the solution}\quad
f^0_a({\bf k},t) = f_{a0}\left[{\bf k}+\frac{e_a}{c}{\bf A}(t)\right],
\label{silin_an}
\end{eqnarray}
where $f_{a0}$ is an arbitrary function depending on the initial conditions.
The equation for $f^1_{a}$ reads
\begin{eqnarray}
\frac{\partial}{\partial t} f^1_a({\bf k},t)+e_a{\bf E}(t)\cdot
\nabla_{\bf k} f^1_a({\bf k},t) = I_a^1({\bf k},t)
\nonumber\\
\mbox{where}\quad
I^1_a({\bf k},t) =
I_{a}\left\{f_a \rightarrow f_{a0}\left[{\bf k}+\frac{e_a}{c}{\bf A}(t)\right]
\right\}.
\label{silin_an1}
\end{eqnarray}
With this scheme, there follow essential simplifications of the above
results because it effectively eliminates the time retardation of the
distribution functions in the collision integrals \cite{sil-an}.
Indeed, one easily verifies that the arguments of the distributions which
appear in the formulas above, now become
\begin{eqnarray}
f_a\left[{\bf k}+{\bf Q}_a(t,t');t'\right] \longrightarrow
f_{a0}\left[{\bf k}+\frac{e_a}{c}{\bf A}(t)\right],
\label{farg-silin}
\end{eqnarray}
and do not depend on the time $t'$ anymore.
This simplification allows to compute the transport, screening and
fluctuation properties quite efficiently.

We first consider the modification of the longitudinal polarization $\Pi^R$.
Straightforward transformations of Eq.~(\ref{pira-gkba}) including a change
of the momentum integration variable lead to
\begin{eqnarray}
\Pi^{R}_{aa}({\bf q};t,t') &=&
e^{i\frac{e_a {\bf q}}{m_a c\hbar}
\int\limits_{t'}^{t}d{\bar t}{\bf A}({\bar t})} \;\cdot
\Pi^{R}_{aa,0}({\bf q};t-t')
\label{pira-silin}
\\
\mbox{whith}\quad
\Pi^{R}_{aa,0}({\bf q};\tau) &=& -\frac{i}{\hbar}\Theta(\tau)\,
\int \frac{d^3 k}{(2\pi\hbar)^3}\,
e^{-\frac{i}{\hbar}
\left(\epsilon^a_{{\bf k+q}}-\epsilon^a_{{\bf k}}\right)\tau
}\,
\left\{f_a({\bf k})-f_a({\bf k+q})\right\}.
\label{pir0a}
\end{eqnarray}
Using this result and the adiabatic approximation (neglecting
the ion contribution to the polarization), we obtain from the Dyson equation,
Eq.~(\ref{vdyson}), for the retarded screened potential
\begin{eqnarray}
V^{R}_{ab}({\bf q};t,t') &=&
e^{i\frac{e_a {\bf q}}{m_a c\hbar}
\int\limits_{t'}^{t}d{\bar t}{\bf A}({\bar t})} \;\cdot
V^{R}_{ab,0}({\bf q};t-t'),
\label{vra-silin}
\end{eqnarray}
where again, $V^{R}_{ab,0}$ denotes the screened potential in the zero
field limit and without retardation in the distribution functions.
Simplifications are possible also for the collision integrals. In particular,
we obtain for the electron--ion scattering term,
\begin{eqnarray}
&& I_{ei}\left({\bf k} - \frac{e_a}{c}{\bf A}(t)\right)
= \frac{2 n_i}{\hbar^2} \,{\rm Re} \int \frac{d^3 q}{(2\pi\hbar)^3}
\left|V_{ei}(q)\right|^2
\int_{-\infty}^{0}d\tau
e^{-\frac{i}{\hbar}
[\epsilon^{e}_{{\bf k}}-\epsilon^{e}_{{\bf k-q}}]\tau}
\nonumber\\
&& \; \times
\int_{0}^{\infty}d\tau_1\int_{-\infty}^{0}d\tau_2\frac{1}
{\epsilon_{0}^{R}({\bf q},\tau_1)}\frac{1}{\epsilon_{0}^{A}({\bf q},\tau_2)} \,
e^{-\frac{i}{\hbar}\frac{e_e}{m_e c}{\bf q}
\int\limits_{t-\tau-\tau_2}^{t-\tau_1} d{\bar t}\,{\bf A}({\bar t})
}
\left\{f_{e0}({\bf k-q})-f_{e0}({\bf k}) \right\}, \quad
\label{iei-silin}
\end{eqnarray}
where $1/\epsilon_{0}^{R/A}$ are the field--free inverse dielectric
functions. Eq.~(\ref{iei-silin}) is the generalization of
Klimontovich's result \cite{klimontovich75}
to quantum plasmas and.

Finally, we consider the simplifications introduced by the ansatz
(\ref{silin_an1}) to the field fluctuations.
The quantity of central importance is the correlation function of the
temporal and spatial microfield fluctuations
$\langle \delta E \delta E \rangle$, Eq.~(\ref{dede}).
A lengthy but straightforward calculation leads to the following
result
\begin{eqnarray*}
&&\overline{
\langle \delta {\bf E}(t_1) \delta {\bf E}(t_2) \rangle }_{{\bf q}} =
\frac{(4\pi)^2}{2 q^2}
e^{i\frac{e_a {\bf q}}{m_a c\hbar}
\int\limits_{t_2}^{t_1}d{\bar t}{\bf A}({\bar t})} \;
\int \frac{d^3 k}{(2\pi\hbar)^3}
\nonumber\\
&&\quad \;\;\;\Bigg\{
\;e_e^2 \,
\frac{e^{-\frac{i}{\hbar}
[\epsilon^{e}_{{\bf k+q}}-\epsilon^{e}_{{\bf k}}](t_1-t_2)} }
{|\epsilon_{0}({\bf q},\epsilon^{e}_{{\bf k+q}}-\epsilon^{e}_{{\bf k}})|^2}
\left\{f_{e0}({\bf k+q})[1-f_{e0}({\bf k})]+
f_{e0}({\bf k})[1-f_{e0}({\bf k+q})]\right\}
\nonumber\\
&&\qquad + \;e_i^2 \,
e^{-\frac{i}{\hbar}[\epsilon^{i}_{{\bf k+q}}-\epsilon^{i}_{{\bf k}}](t_1-t_2)}
\int_{0}^{\infty} d\tau_1 \,
\frac{e^{-\frac{i}{\hbar}
[\epsilon^{i}_{{\bf k+q}}-\epsilon^{i}_{{\bf k}}]\tau_1} }
{\epsilon_{0}^{R}({\bf q},\tau_1)}
\nonumber\\
&&\qquad\quad\times
\int^{0}_{-\infty} d\tau_2 \,
\frac{e^{\frac{i}{\hbar}
[\epsilon^{i}_{{\bf k+q}}-\epsilon^{i}_{{\bf k}}]\tau_2} }
{\epsilon_{0}^{A}({\bf q},\tau_2)} \;
e^{-\frac{i}{\hbar}\left(\frac{e_e}{m_e}-\frac{e_i}{m_i}\right){\bf q}
\int\limits_{t_2-\tau_2}^{t_1-\tau_1} d{\bar t}\,\frac{{\bf A}}{c}({\bar t})
}
\left[f_{i0}({\bf k+q})+f_{i0}({\bf k})\right]
\Bigg\},
\nonumber
\label{de-silin}
\end{eqnarray*}
where the first term in parantheses (second line) is the electron
contribution, and the second (third and fourth line) results from the
ions. Again, this is a generalization of Klimontovich's remarkable
result \cite{klimontovich75} who considered the classical limit and
the equal time fluctuations, $t_1=t_2$. Our result fully includes
the two--time fluctuations which are directly measurable quantities.
From the above fluctuation spectrum, all major observables of dense
quantum plasmas in a strong laser field can be computed. The corresponding
analysis will be presented in a forthcoming paper.

\section{Discussion}
In this paper, we have presented a gauge--invariant derivation of the
quantum kinetic equation for dense plasmas in a laser field. Our main
result, Eq.~(\ref{coli3}), generalizes previous work to quantum
systems. This equation can be used to calculate the transport properties
of a dense plasma in a laser field on arbitrary time scales, i.e., over
the whole frequency range.
The use of the random phase approximation allows for a highly consistent
treatment of the combined effect of internal longitudinal fields
(dynamical screening) and transverse electromagnetic fields, including
intense laser pulses. In particular, it allows to investigate the influence
of the electromagnetic field on the two--particle scattering process and
the screening properties of the plasma and on the screening buildup
in the presence of a strong field.

Besides, the presented gauge--invariant approach is completely general and
can be extended straightforwardly to more complex situations, including
strong coupling effects, bound states, impact and field ionization.
Moreover, it can be directly
generalized to relativistic systems und ultra--intense fields.


\section*{Acknowledgments}
It is a great pleasure to dedicate this paper to Yuri Lvovich
Klimontovich on his seventyfith birthday. All of us have, at
different times, come in contact with the remarkable papers and
monographs of Klimontovich. We have very much enjoyed countless
discussions with him as well as rewarding scientific collaborations
on problems of kinetic theory of nonideal gases and plasmas in
general, in particular on dielectric properties \cite{KK74},
kinetic equations with bound states \cite{KK81,KKK87}, foundation
of reaction--diffusion equations for nonideal plasmas \cite{KSB88},
selforganization and structure formation \cite{KB88} and many others.

The authors acknowledge stimulating discussions with G. Kalman, P. Mulser
and R. Sauerbrey.
This work is supported by the Deutsche Forschungsgemeinschaft
(Schwerpunkt ``Wechselwirkung intensiver Laserfelder mit Materie''),
and by the European Commission through the TMR Network SILASI.




%
\begin{received}Received
\end{received}

\begin{thebibliography}{10}
\bibitem{n-klim}
It is interesting to note that Klimontovich's phase space densities
$N$ \cite{Kli57} are essentially the classical limit of the field
operators. Therefore, his kinetic theory is of similar generality
as modern field theory, cf. his text books
\cite{klimontovich64,klimontovich75,klimontovich82}.
\bibitem{sil-an} This ansatz is analogous to
the zeroth order retardation ansatz of non--Markovian field--free
kinetic equations, e.g. \cite{bonitz-book}.
\end{thebibliography}
\end{document}